\documentclass[3p,times,procedia]{elsarticle}
\usepackage{nupha_ecrc}
\usepackage{hyperref}

\volume{00}

\firstpage{1}

\journalname{Nuclear Physics A}

\runauth{}

\jid{nupha}

\jnltitlelogo{Nuclear Physics A}

\usepackage{amssymb}
 \usepackage{amsthm}
 \usepackage{amsmath}

\usepackage{multirow}
\usepackage{tabularx}
\usepackage{changepage}     
\usepackage{lineno}   
\usepackage[figuresright]{rotating}

\newcommand{\pT}{$p_{\mathrm{T}}$}

\newcommand{\sNN}{$\sqrt{s_{\mathrm{NN}}}=5.02$ TeV}

\makeatletter
\newcommand{\mathleft}{\@fleqntrue\@mathmargin0pt}
\newcommand{\mathcenter}{\@fleqnfalse}
\makeatother

\begin{document}

\begin{frontmatter}

 \title{Non-linear flow modes of identified particles in Pb--Pb collisions at $\sqrt{s_{\mathrm{NN}}} = 5.02$ TeV with the ALICE detector}
 \author{Naghmeh Mohammadi for the ALICE collaboration}
 
 \ead{naghmeh.mohammadi@cern.ch}
 
 \address{CERN, Route de Meyrin, 1211 Geneva, Switzerland}
\dochead{XXVIIth International Conference on Ultrarelativistic Nucleus-Nucleus Collisions\\ (Quark Matter 2018)}

\title{Non-linear flow modes of identified particles in Pb--Pb collisions at $\sqrt{s_{\mathrm{NN}}} = 5.02$ TeV with the ALICE detector}
\begin{abstract}
\noindent The non-linear flow modes, $v_{4,22}$, $v_{5,32}$ and $v_{6,33}$ as a function of transverse momentum ($p_{\mathrm{T}}$) have been measured for the first time in Pb--Pb collisions at $\sqrt{s_{\rm{NN}}} = 5.02$ TeV for identified particles, i.e. charged pions, kaons and (anti-)protons. These observables probe the contribution of the second and third initial anisotropy coefficients to the higher order flow harmonics. Interestingly, all the characteristic features observed in previous $p_{\mathrm{T}}$-differential as well as $p_{\mathrm{T}}$-integrated measurements (e.g. $v_{2}$ and $v_{3}$) for various particle species are present in these measurements, i.e. increase of magnitude with increasing centrality percentile, a mass ordering in the low $p_{\mathrm{T}}$ region and a particle type grouping in the intermediate $p_{\mathrm{T}}$ region. The results cover 0--1\% (ultra-central collisions), 10--20\% (mid-central collisions) and 40--50\% (mid-peripheral collisions) centrality intervals and allow to test models that attempt to describe initial conditions and the mode-coupling of different harmonics. 
\end{abstract}

\end{frontmatter}

\section{Introduction}
\label{sec:Intro}

\vspace{-0.15cm}

The goal of heavy-ion collision experiments is to study the properties of a state of matter that exists at extremely high temperatures and energy densities, namely, the quark-gluon plasma. One of the main observables to probe the properties of this state of matter is anisotropic flow. It arises from the initial spatial anisotropy caused by the geometry of the collision together with initial-state fluctuations of the participating nucleons in the collision. This initial spatial anisotropy transfers through the low viscous medium into the momentum anisotropy of the final state particles. As a result, anisotropic flow can probe the initial conditions, the equation of state and the transport properties of the medium, e.g. specific shear viscosity ($\eta/s$) and bulk viscosity ($\zeta/s$). The magnitude of the anisotropic flow is quantified by the coefficients ($v_{\mathrm{n}}$) of the Fourier expansion of the azimuthal distribution of particle production relative to the symmetry plane, according to \cite{Voloshin:1994mz}: 
\vspace{-0.1cm}
\begin{equation}
\frac{\mathrm{d}N}{\mathrm{d}(\varphi - \Psi_{\mathrm{n}})} \approx 1+2\sum_{\rm{n}=1}^{\infty}v_{\mathrm{n}}\cos[\mathrm{n}(\varphi - \Psi_{\mathrm{n}})],
\end{equation}

\noindent where $\varphi$ is the azimuthal angle and $\Psi_{\mathrm{n}}$ the n-th order symmetry plane angle. 

It has been shown multiple times that the lower order flow vectors, $V_{2}$ and $V_{3}$, are linearly correlated with the same order initial spatial anisotropy ($\varepsilon_{\rm{n}}$) \cite{Alver:2010gr}. However, higher harmonics ($\rm{n}>3$) have contributions from not only their corresponding spatial anisotropy, but also the lower order spatial anisotropies, $\varepsilon_{2}$ and/or $\varepsilon_{3}$ \cite{Bhalerao:2014xra}. In fact one could decompose higher order flow vectors into linear and non-linear response. For instance the fourth, fifth and sixth order flow vectors can be written according to \cite{Bhalerao:2014xra}: 

\vspace{-0.55cm}
\begin{align}
V_{4} &= V_{4}^{\mathrm{L}} + V_{4}^{\mathrm{NL}} = V_{4}^{\mathrm{L}} + \chi_{4,22}(V_{2})^2, \nonumber \\
V_{5} &= V_{5}^{\mathrm{L}} + V_{5}^{\mathrm{NL}} = V_{5}^{\mathrm{L}} + \chi_{5,32}V_{3}V_{2}, \nonumber \\
V_{6} &= V_{6}^{\mathrm{L}} + V_{6}^{\mathrm{NL}} = V_{6}^{\mathrm{L}} + \chi_{6,222}(V_{2})^3 + \chi_{6,33}(V_{3})^2 + \chi_{6,42}V_{2}V_{4}^{\mathrm{L}},
\label{Eq:V4V5V6}
\end{align}
\vspace{-0.55cm}

\noindent where $V_{\rm{n}}^{\mathrm{L}}$ and $V_{\rm{n}}^{\mathrm{NL}}$ are the linear and non-linear flow vectors which are proven to be uncorrelated \cite{Acharya:2017zfg}. The magnitude of these flow vectors are denoted as $v_{\rm{n}}^{\mathrm{L}}$ for the linear mode and $v_{\rm{n,mk}}$ for the non-linear mode, where $\rm{n}$ stands for the order of the higher harmonic flow and $\rm{m}$ and $\rm{k}$ stand for the lower order anisotropic flow. Using a generic framework \cite{Bilandzic:2013kga} updated with an implementation of the sub-event method the magnitude of $v_{4,22}$, $v_{5,32}$ and $v_{6,33}$ can be extracted via \cite{Acharya:2017zfg}:
\vspace{-0.45cm}

\begin{align}
v_{4,22}^{\rm{A}}(p_{\rm{T}}) &= \frac{\langle v_{4}(p_{\rm{T}})v_{2}^{2}\cos(4\Psi_{4}-4\Psi_{2})\rangle}{\sqrt{\langle v_{2}^{4}\rangle}} = \frac{\langle\langle \cos(4\varphi^{\rm{A}}_{1}(p_{\rm{T}})-2\varphi^{\rm{\rm{B}}}_{2}-2\varphi^{\rm{B}}_{3})\rangle\rangle}{\sqrt{\langle\langle \cos(2\varphi^{\rm{A}}_{1}+2\varphi^{\rm{A}}_{2}-2\varphi^{\rm{B}}_{3}-2\varphi^{\rm{B}}_{4}) \rangle\rangle}}, \nonumber \\
v_{5,32}^{\rm{A}}(p_{\rm{T}}) &= \frac{\langle v_{5}(p_{\rm{T}})v_{2}v_{3}\cos(5\Psi_{5}-2\Psi_{2}-3\Psi_{3})\rangle}{\sqrt{\langle v_{2}^{2}v_{3}^{2}\rangle}} = \frac{\langle\langle \cos(5\varphi^{\rm{A}}_{1}(p_{\rm{T}})-2\varphi^{\rm{B}}_{2}-3\varphi^{\rm{B}}_{3})\rangle\rangle}{\sqrt{\langle\langle \cos(2\varphi^{\rm{A}}_{1}+3\varphi^{\rm{A}}_{2}-2\varphi^{\rm{B}}_{3}-3\varphi^{\rm{B}}_{4}) \rangle\rangle}}, \nonumber \\
v_{6,33}^{\rm{A}}(p_{\rm{T}}) &= \frac{\langle v_{6}(p_{\rm{T}})v_{3}^{2}\cos(6\Psi_{6}-6\Psi_{3})\rangle}{\sqrt{\langle v_{3}^{4}\rangle}} = \frac{\langle\langle \cos(6\varphi^{\rm{A}}_{1}(p_{\rm{T}})-3\varphi^{\rm{B}}_{2}-3\varphi^{\rm{B}}_{3})\rangle\rangle}{\sqrt{\langle\langle \cos(3\varphi^{\rm{A}}_{1}+3\varphi^{\rm{A}}_{2}-3\varphi^{\rm{B}}_{3}-3\varphi^{\rm{B}}_{4}) \rangle\rangle}},
\label{Eq:Vnmk}
\end{align}

\vspace{-0.15cm}

\noindent where $v_{\rm{n,mk}}^{\rm{A (B)}}$ is measured by selecting the particle of interest from subevent ``A" (``B") and the reference particles from subevent ``B" (``A"). The average of $v_{\rm{n,mk}}^{\rm{A}}$ and $v_{\rm{n,mk}}^{\rm{B}}$ quantifies the magnitude of the non-linear flow modes. In these proceedings, we present preliminary measurements of the magnitude of the non-linear flow terms for the fourth, fifth and the sixth flow harmonics of charged pions, kaons and (anti-)protons.  The study of the $p_{\rm{T}}$-differential non-linear flow modes of identified particles could yield more information about the particle production mechanisms, such as quark coalescence, as well as the hadronic rescattering phase. 

\vspace{-0.45cm}
\section{Analysis details}
\vspace{-0.15cm}

The analysis is done on the minimum-bias Pb--Pb data at \sNN~collected by ALICE \cite{Aamodt:2008zz} in 2015. About 45 million events are analysed in total with 0.5 million events per 1\% centrality bin. The minimum bias Pb--Pb events are triggered by the coincidence between the signals from both sides of the V0 detector. The track reconstruction is performed using the information from both the Time Projection Chamber (TPC) and the Inner Tracking System (ITS). The momenta of charged particles are measured using both ITS and TPC with a pseudorapidity range of $|\eta|<0.9$. The particle identification (PID) for pions ($\pi^{\pm}$), kaons ($K^{\pm}$) and (anti-)protons ($p+\bar{p}$) in this analysis is 
based on a $p_{\rm{T}}$-dependent approach which combines the signals from the TPC and Time-Of-Flight (TOF) detectors. In this analysis, PID for charged pions maintains a minimum purity of 90\% up to $p_{\rm{T}} = 6$ GeV$/c$, for charged kaons a minimum purity of 80\% up to $p_{\rm{T}} = 4$ GeV$/c$ and finally for (anti-)protons a minimum purity of 80\% up to $p_{\rm{T}} = 6$ GeV$/c$. 

 The $v_{\rm{n,mk}}$ measurements are performed using a two sub-event generic framework method where identified hadrons and reference charged particles are selected from two non-overlapping sub-events in pseudo-rapidity. Using multi-particle correlations reduces the non-flow contributions, i.e. correlations unrelated to the common symmetry plane. The residual non-flow effects have been investigated by requiring a larger $\eta$ gap between the sub-events. 

\vspace{-0.43cm}
\section{Results and discussion}
\vspace{-0.15cm}

\begin{figure}[htb]
\begin{center}
\includegraphics[scale=0.2286]{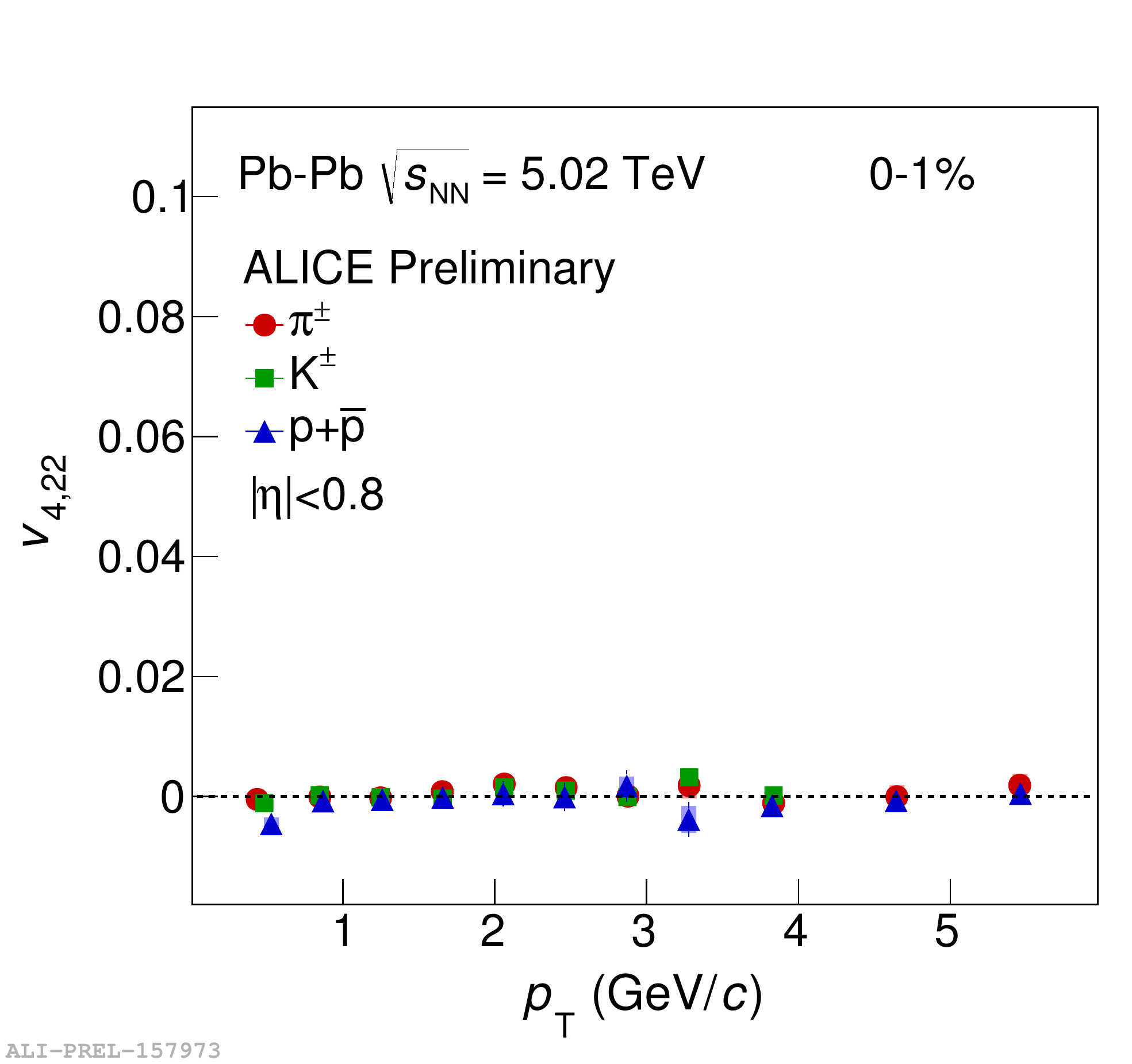} 
\includegraphics[scale=0.2286]{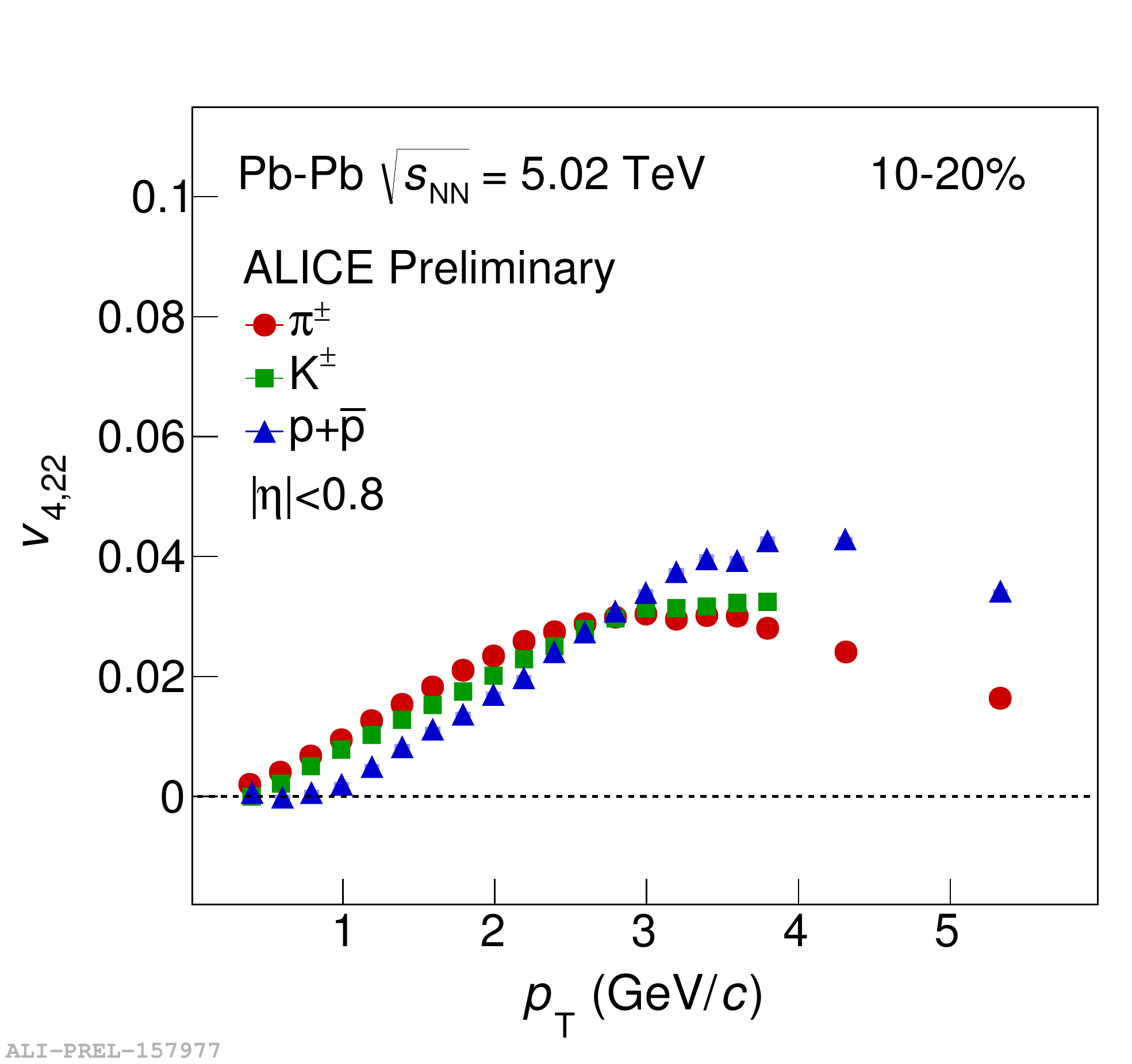}
\vspace{-0.055cm}
\includegraphics[scale=0.2286]{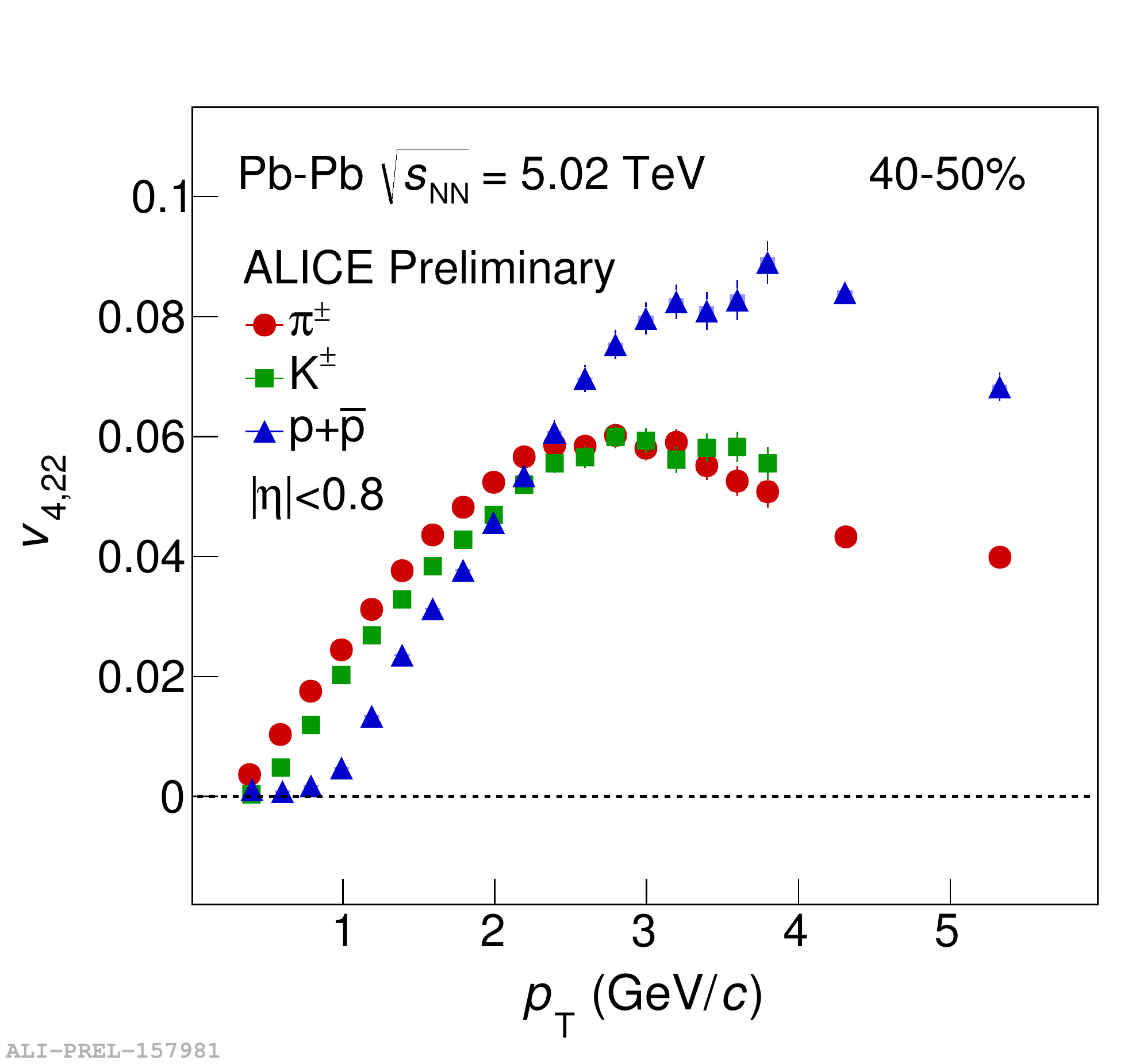}
\includegraphics[scale=0.2286]{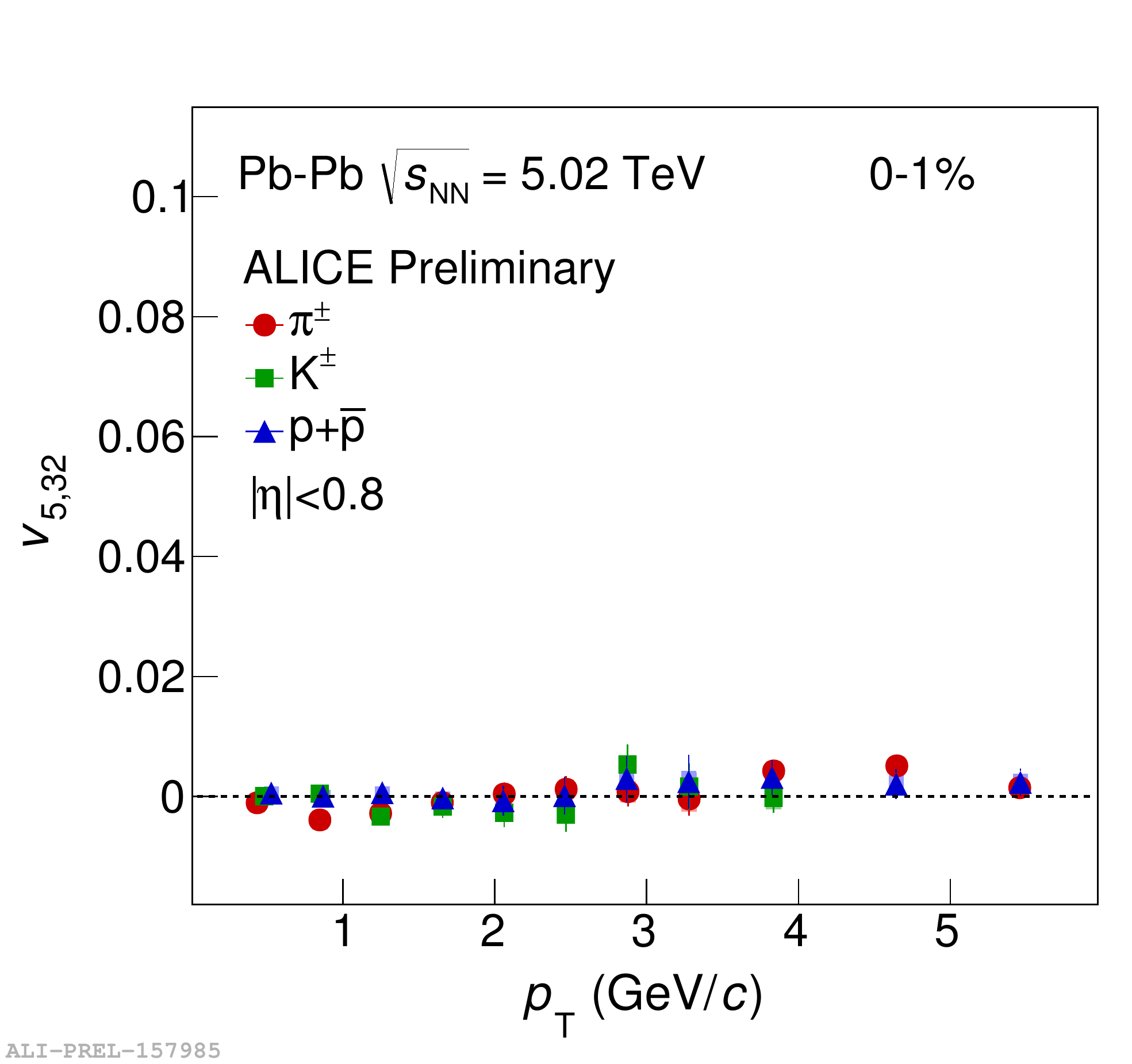}
\vspace{-0.062cm}
\includegraphics[scale=0.2286]{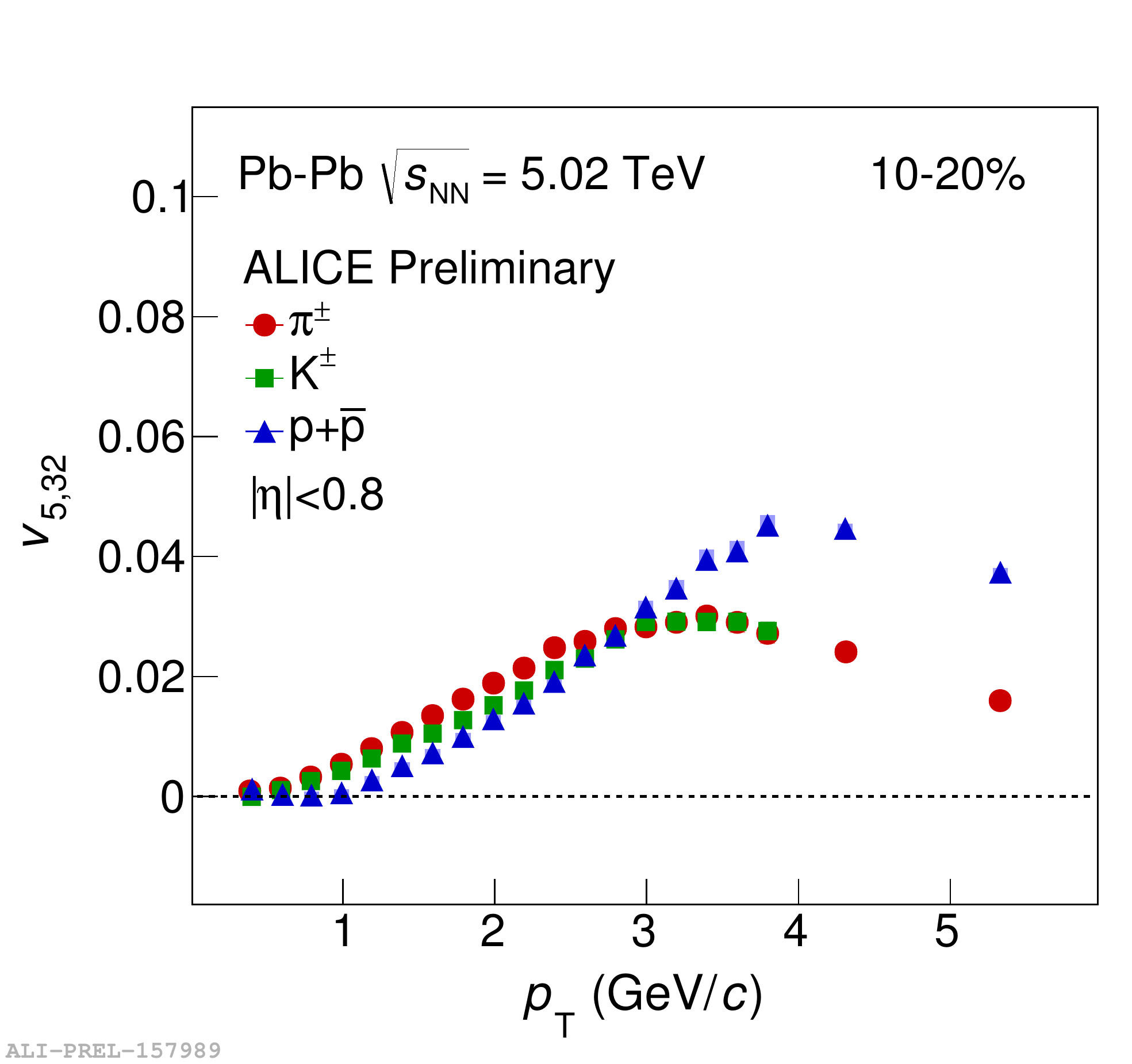}
\includegraphics[scale=0.2286]{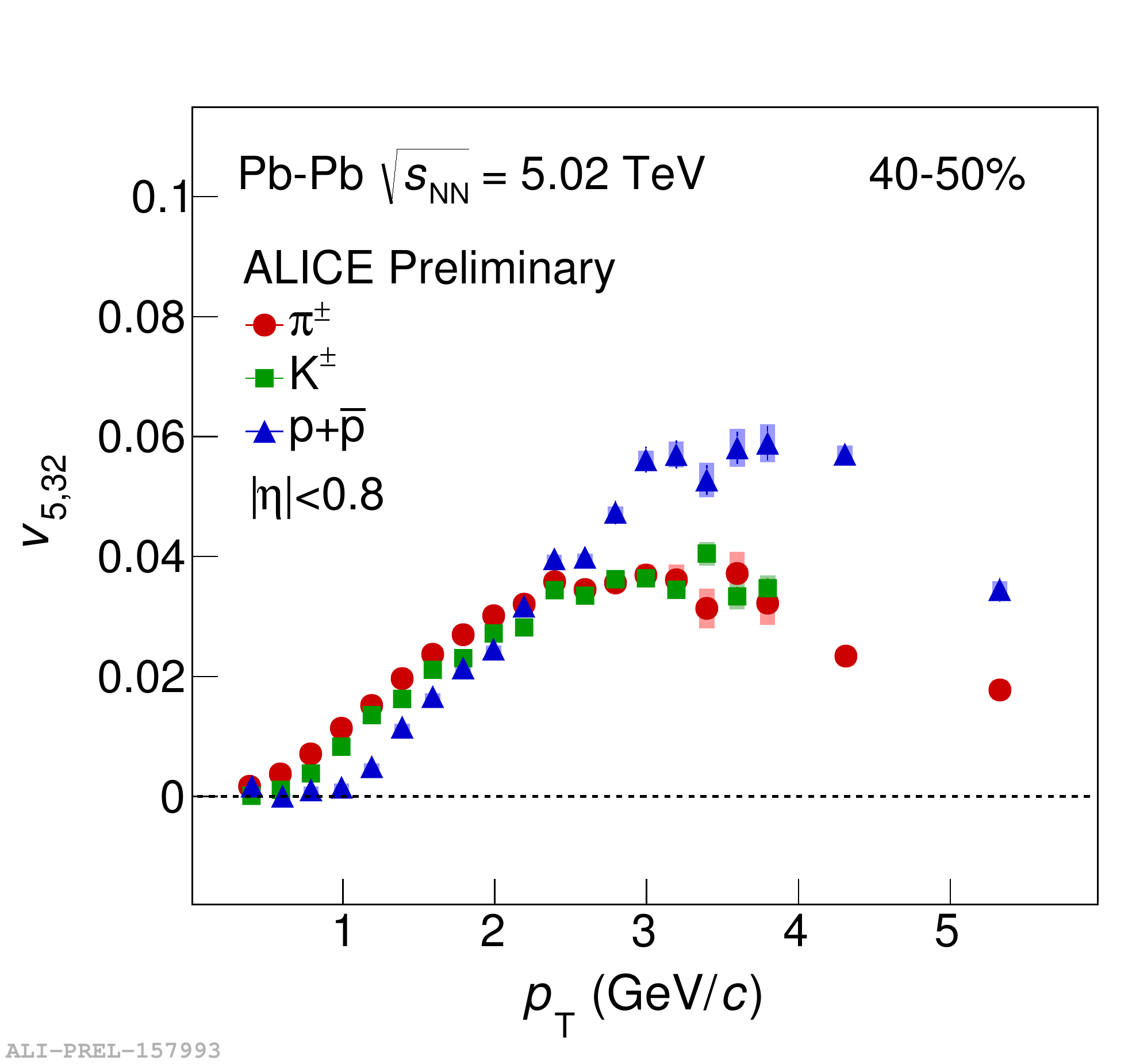}
\includegraphics[scale=0.2286]{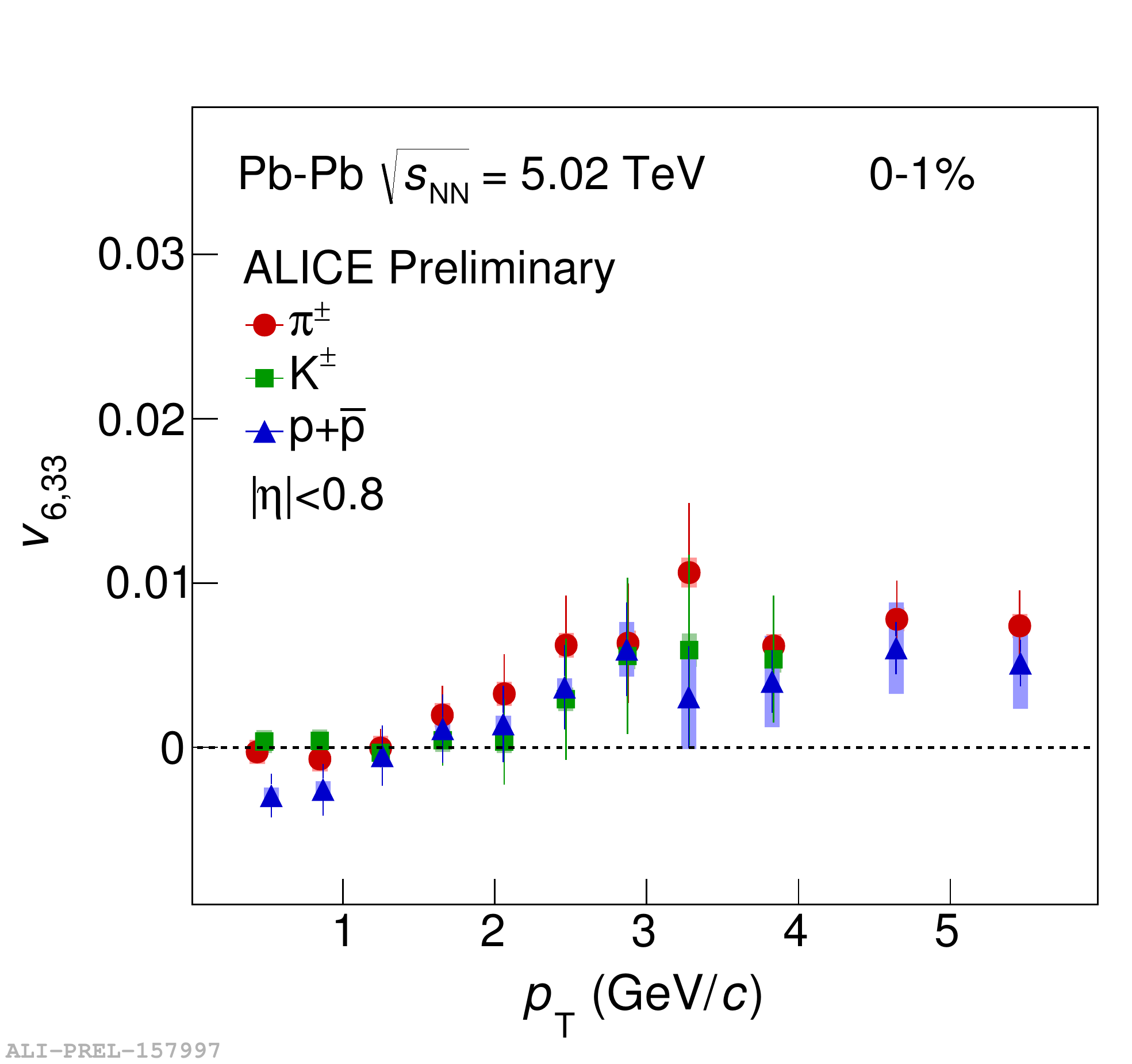}
\includegraphics[scale=0.2286]{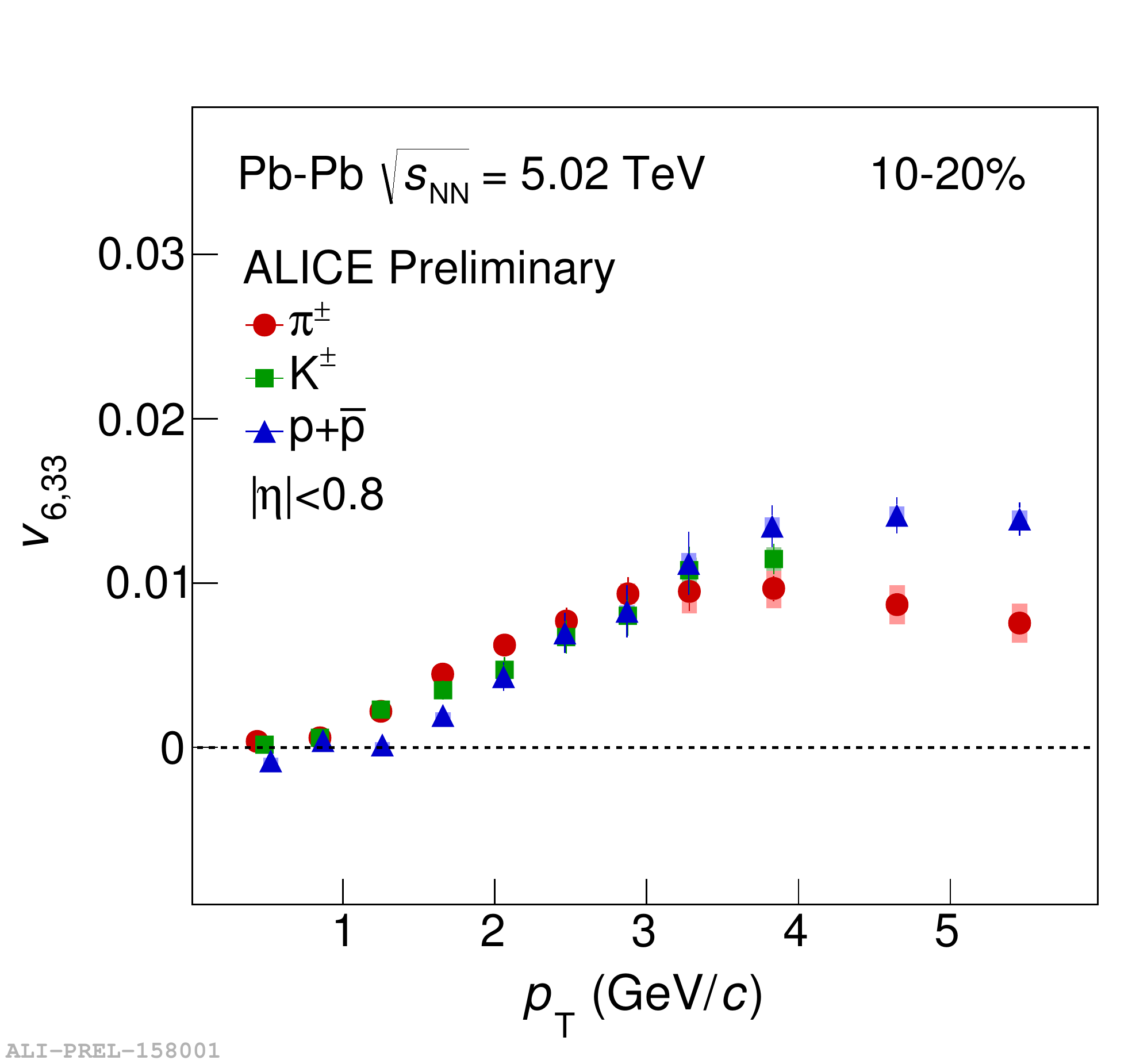}
\includegraphics[scale=0.2286]{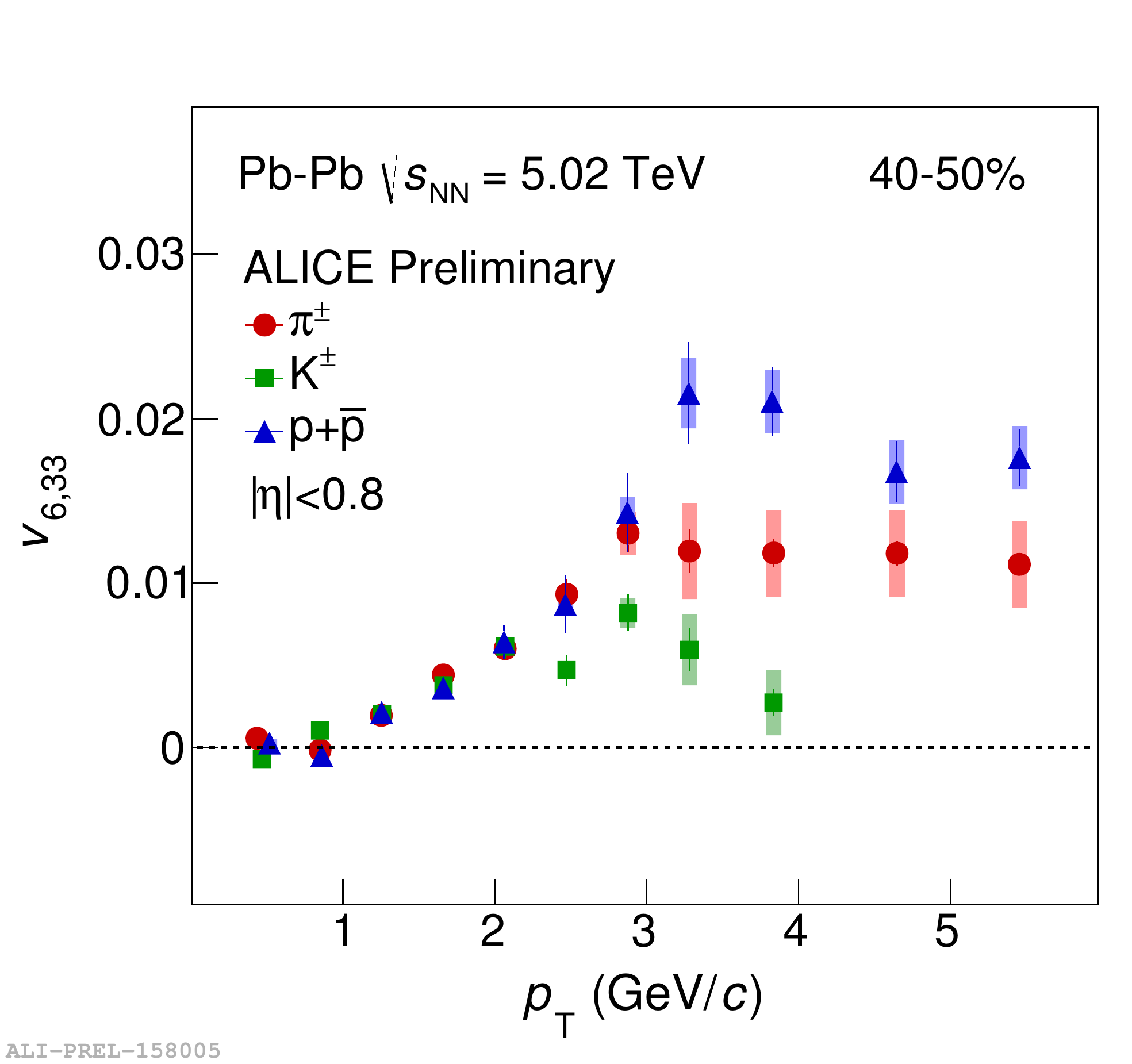}
\end{center}
\vspace{-0.2cm}
\caption{The \pT-differential $v_{4,22}$ (top row), $v_{5,32}$ (middle row) and $v_{6,33}$ (bottom row) for different particle species grouped into different centrality intervals of Pb--Pb collisions at\sNN.}
\label{vnmk_particleDependence}
\end{figure}

Figure \ref{vnmk_particleDependence} presents the \pT-differential $v_{4,22}$ (top row), $v_{5,32}$ (middle row) and $v_{6,33}$ (bottom row), for charged pions, kaons and (anti-)protons. Figures are grouped into 0--1\% (left column - ultra-central collisions), 10--20\% (middle column - semi-central collisions) and 40--50\% (right column - mid-peripheral collisions) centrality intervals of Pb--Pb collisions at \sNN. In the ultra-central collisions where the evolution of the system is mainly driven by the initial state fluctuations, the magnitude of $v_{4,22}$ and $v_{5,32}$ remain zero for all particle species and $p_{\rm{\rm{T}}}$ values. The coefficient $v_{6,33}$ exhibits a non-zero magnitude which could be due to higher sensitivity to the initial state fluctuations. By increasing the centrality percentile, the magnitude of the \pT-differential $v_{4,22}$ and $v_{5,32}$ rise steeply while the magnitude of $v_{6,33}$ exhibits a milder increase. Furthermore, a clear mass ordering can be seen in the low \pT~region (i.e. $p_{\rm{T}}\leq 2.5$ GeV$/c$) at semi-central and mid-peripheral collisions for $v_{4,22}$ and $v_{5,32}$, which arises from the interplay between the non-linear response of the system and radial flow. Radial flow creates a depletion in the particle spectra at lower \pT~values which leads to lower $v_{\rm{n}}$ and in turn, lower $v_{\rm{n,mk}}$ for heavier particles \cite{Adam:2016nfo}. In the intermediate \pT~region (\pT$\geq 2.5$ GeV$/c$) of $v_{4,22}$ and $v_{5,32}$, a particle type grouping can be seen for mesons ($\pi^{\pm}$ and $K^{\pm}$) and a clear separation of the baryons ($\rm{p}+\bar{\rm{p}}$) is observed. In addition, the $v_{\rm{n,mk}}$ values for baryons are larger than for mesons.
\vspace{-0.4cm}

\vspace{-0.15cm}

A comparison is performed between the measured non-linear flow modes and two hydrodynamical calculations from \cite{Zhao:2017yhj}. Both of these calculations are based on iEBE-VISHNU, an event-by-event version of the VISHNU hybrid model coupling $2+1$ dimensional viscous hydrodynamics (VISH$2+1$) to a hadronic cascade model (UrQMD). The first model uses AMPT initial conditions with constant values of specific shear viscosity ($0.08$, the lower limit conjectured by AdS/CFT) and bulk viscosity ($\zeta/s = 0$), and the second model incorporates TRENTo \cite{Bernhard:2016tnd} initial conditions with a temperature dependent specific shear and bulk viscosity. For simplicity in the rest of these proceedings, the first model is referred to as AMPT and the second model as TRENTo.

\begin{figure}[htb]
\begin{center}
\includegraphics[scale=0.27]{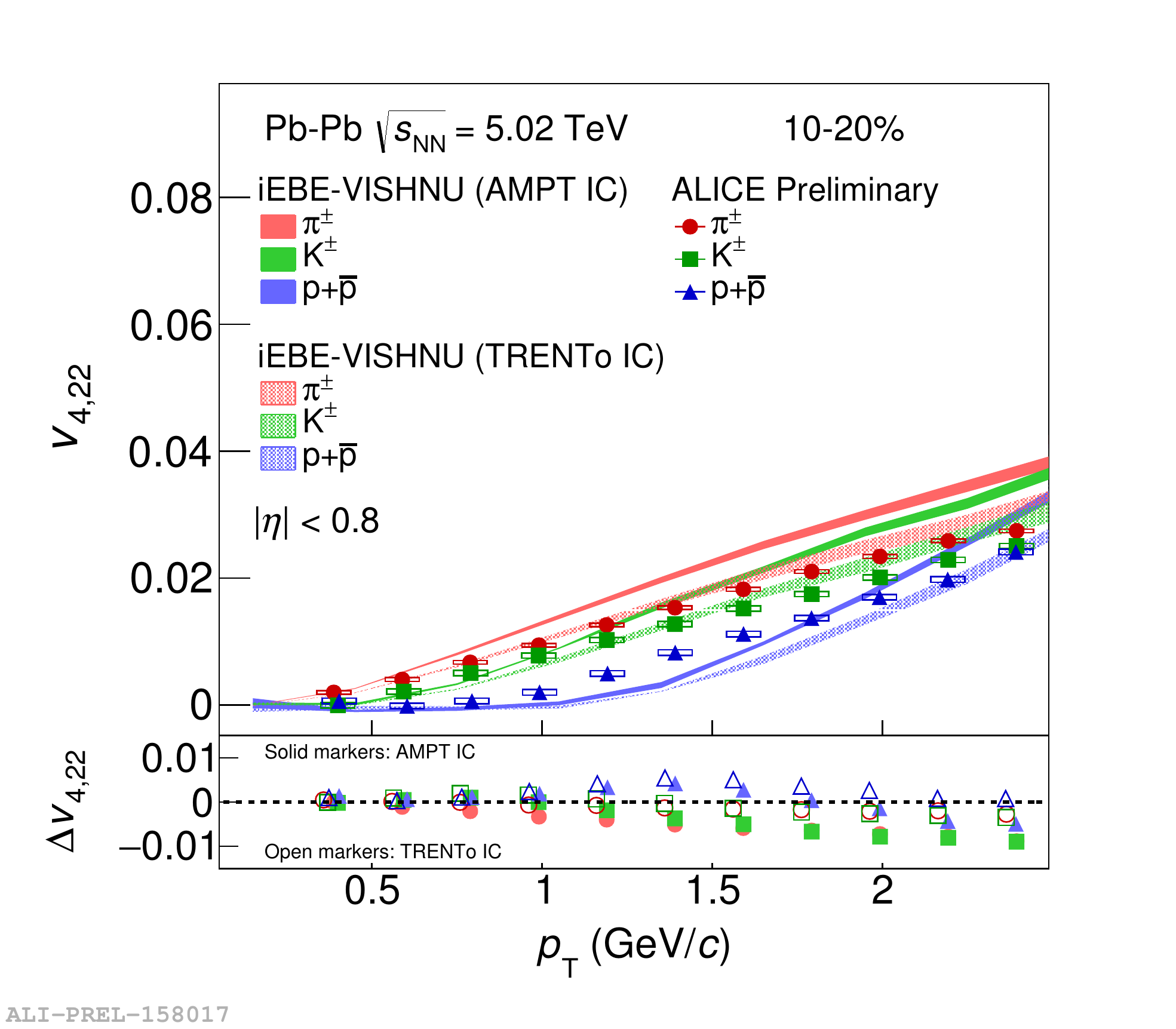} \hspace{-0.75cm}
\includegraphics[scale=0.27]{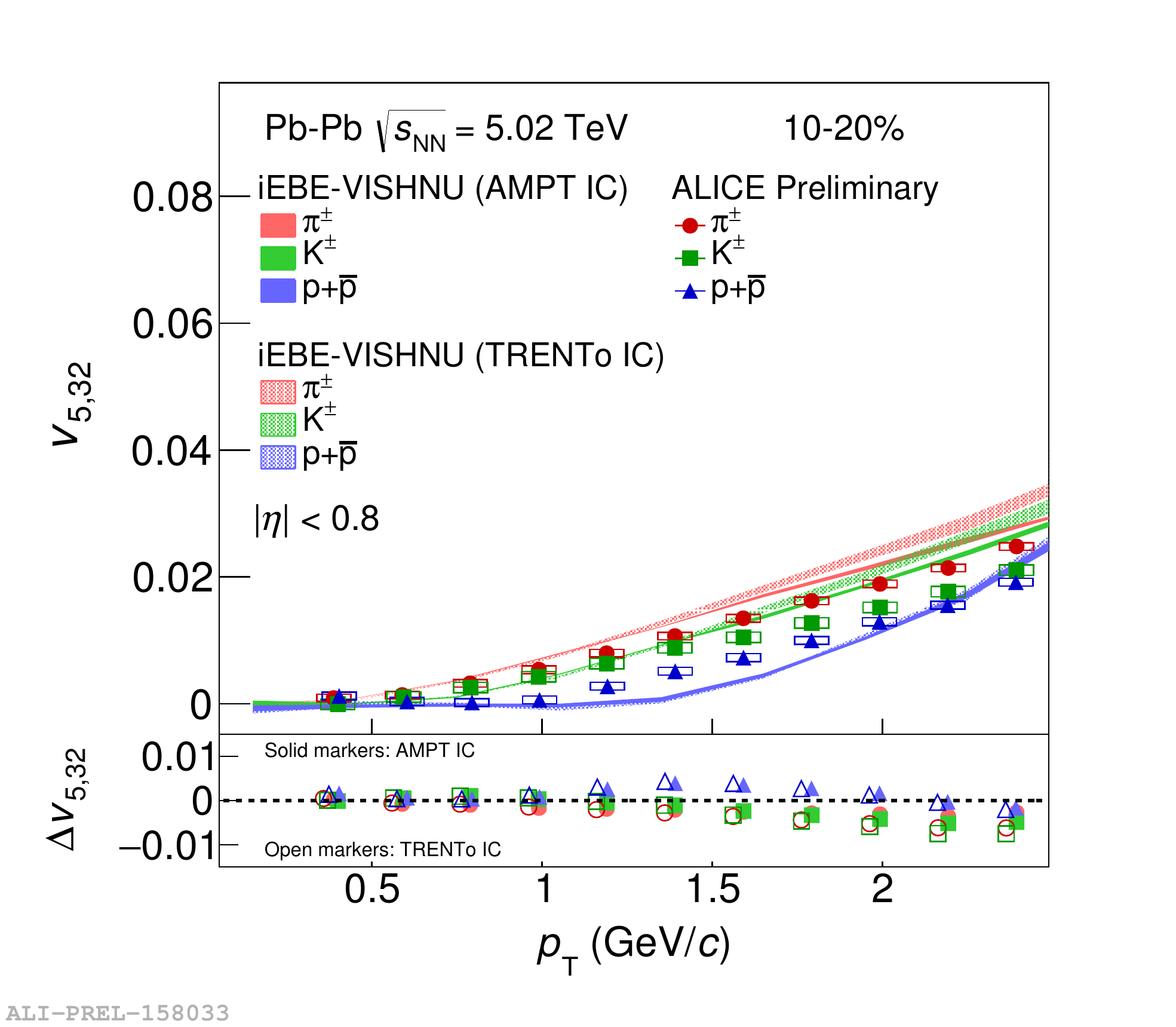} \hspace{-0.75cm}
\includegraphics[scale=0.27]{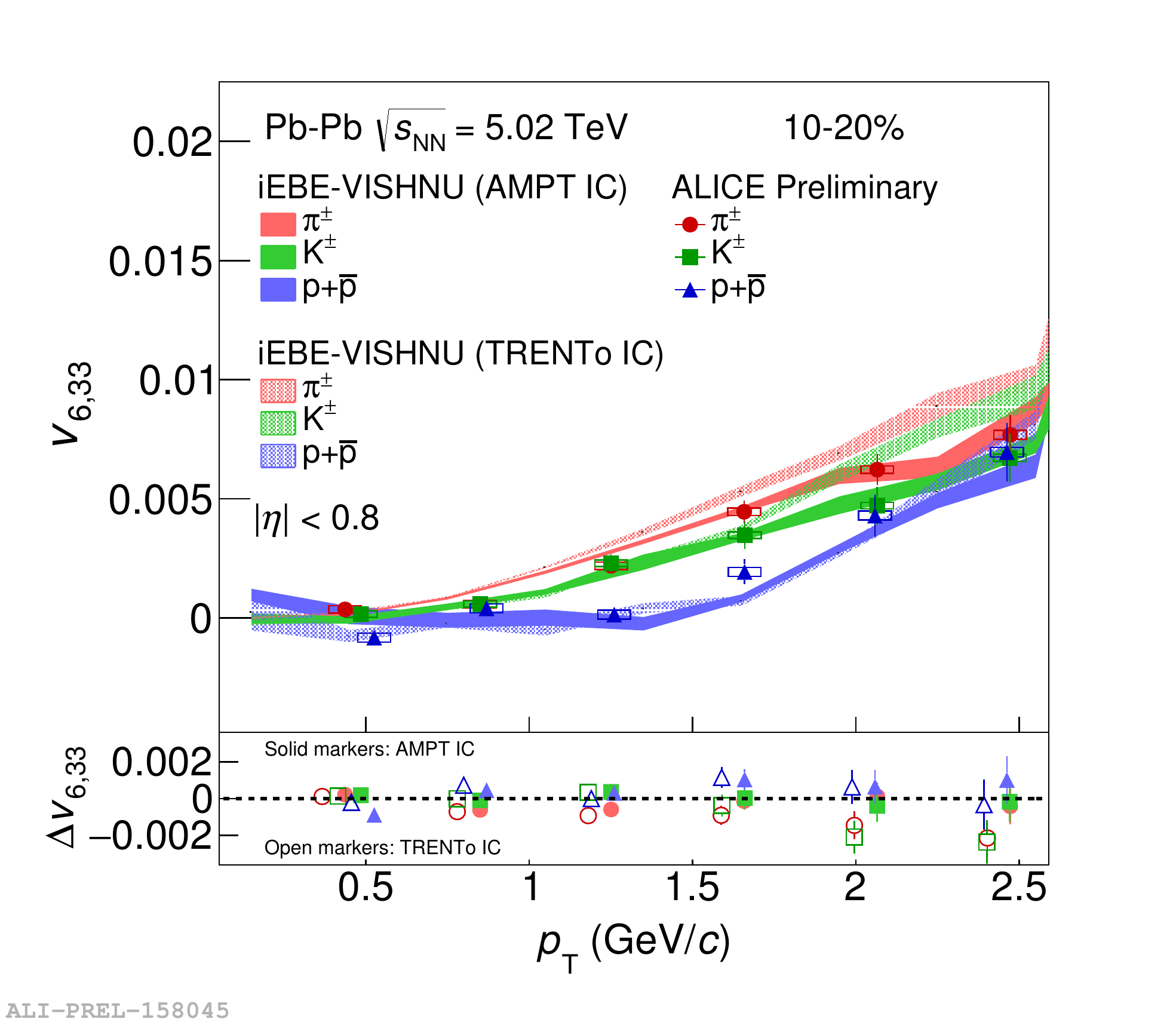}

\end{center}
\vspace{-0.45cm}
\caption{The \pT-differential $v_{4,22}$ (left column), $v_{5,32}$ (middle column) and $v_{6,33}$ (right column) for different particle species in 10--20\% centrality intervals of Pb--Pb collisions at\sNN~compared with iEBE-VISHNU hybrid models with two different sets of initial parameters: AMPT initial conditions ($\eta/s = 0.08$ and $\zeta/s = 0$) shown in solid bands and TRENTo initial conditions ($\eta/s(\rm{T})$ and $\zeta/s(\rm{T})$) in hatched bands. The bottom panels show the difference between the measurements and each model.}
\label{vnmk_comparison}
\end{figure}

Figure \ref{vnmk_comparison} presents the comparison between the measurements and both models for the \pT-differential $v_{4,22}$ (left), $v_{5,32}$ (middle) and $v_{6,33}$ (right) for charged pions, kaons and (anti-)protons at 10--20\% centrality interval. The solid bands show the AMPT model and the hatched bands represent the TRENTo calculations. The bottom panels in each plot in Fig. \ref{vnmk_comparison} present the difference between the models and the measurement. Both models produce a mass ordering in $p_{\rm{T}}\leq 2.5$ GeV$/c$. However, $v_{4,22}$ and $v_{5,32}$ are only qualitatively described by the models. Finally, AMPT is in slightly better agreement with $v_{6,33}$ measurements.
\vspace{-0.38cm}

\vspace{-0.8cm}
\section{Summary}
\vspace{-0.15cm}

In these proceedings, the preliminary measurements of the $p_{\mathrm{T}}$-differential non-linear flow modes, $v_{4,22}$, $v_{5,32}$ and $v_{6,33}$ have been shown for charged pions, kaons and (anti-)protons for 0--1\%, 10--20\% and 40--50\% centrality intervals in Pb--Pb collisions at \sNN. The magnitude of all non-linear flow modes exhibit a centrality dependence. This centrality dependence is much larger for $v_{4,22}$ and $v_{5,32}$ which originates from the contribution of second order flow harmonic, as shown in Eq. \ref{Eq:V4V5V6}, and reflects the dependence of $v_{2}$ on the anisotropy of the collision geometry. $v_{6,33}$ does not exhibit a considerable centrality dependence since it is mainly sensitive to the initial state fluctuations and less driven by the collision geometry. A clear mass ordering is seen in the low \pT~region (\pT$<2.5$ GeV$/c$) as well as a particle type grouping in the intermediate \pT~region (\pT$>2.5$ GeV$/c$) at both 10--20\% and 40--50\% centrality intervals. The comparison of two models based on the iEBE-VISHNU hybrid model, but with two different initial conditions (AMPT and TRENTo) and transport properties show that neither of the models are able to fully describe the measurements. The model using AMPT initial conditions ($\eta/s = 0.08$ and $\zeta/s = 0$) exhibits a magnitude and shape closer to the measurements. As a result, in order to further constrain the values of transport properties and the initial conditions of the system, it is necessary to tune the input parameters of future hydrodynamic calculations attempting to describe these measurements.

\bibliographystyle{elsarticle-num}

\end{document}